\documentstyle[prl,twocolumn,aps,english]{revtex}

\def\6{\langle}
\def\9{\rangle}
\def\tr{{\rm tr}}

\begin{document}

\title{What's Wrong with these Observables?}

\author{Asher Peres}
\address{Department of Physics, Technion---Israel Institute of
Technology, 32000 Haifa, Israel}

\maketitle
\begin{abstract}
An imprecise measurement of a dynamical variable (such as a spin
component) does not, in general, give the value of another dynamical
variable (such as a spin component along a slightly different
direction). The result of the measurement cannot be interpreted as the
value of any observable that has a classical analogue.
\end{abstract}

\bigskip

\noindent{\bf1. MEMORABILIA}\bigskip

\noindent I was never compelled to learn solid state physics, and my
first encounter with David Mermin was his paper \cite{pt} {\it ``Is the
moon there when nobody looks? Reality and the quantum theory''\/}. That
paper gave a wonderful proof of Bell's theorem. I wrote a comment
\cite{susan} saying ``Mermin gives a remarkably simple illustration of
Bell's theorem, but leaves the impression that something mysterious is
implied. The situation is much simpler: the pair of photons is a single,
nonlocal, indivisible entity \ldots\ It is only because we force upon
the photon pair the description of two separate particles that we get
the paradox of Einstein, Podolsky and Rosen.'' David answered to me,
privately, ``one person's mystery is another person's explanation.''

The following year, we met in a conference that Danny Greenberger had
organized in the World Trade Center in New York City \cite{nyas} and
it was friendship at first sight. I gave a talk ``When is a quantum
measurement?'' after which Abner Shimony came to me and said, with
a big friendly smile ``Asher, you understand nothing! You speak just
like Niels Bohr!'' One person's insult is another person's compliment.

Over the years, David and I collaborated many times, but I never had
the honor of being his co-author. Many of our discussions revolved
around the Kochen-Specker (KS) theorem~\cite{ks}. In 1990, David
relayed to me information that Conway and Kochen had found a new proof
of that theorem using only 33 rays, instead of 117 in the original
proof. The only clue I had was that they used a cubic lattice for
constructing their rays. This led me to find a much simpler proof, also
with 33 rays, but this one using an {\it irrational\/} lattice. I wrote
to Simon Kochen, asking him to inform me when he publishes his proof,
so that I could publish mine immediately afterwards. Kochen answered
that my proof was indeed very simple, but meanwhile he had found one
with only 31 rays. I was both happy and disappointed. However, soon
afterwards, I found a proof with only 24 rays, in four dimensions. I
wrote to Kochen that $24<31$ and that I would publish my results,
while mentioning the existence of their proof with 31 rays
\cite{jpa}. Conway and Kochen then never bothered to publish the latter.

In my paper \cite{jpa} I also stated that 24 was the minimal number
number in four dimensions. David was not convinced by my argument.
We exchanged several letters on this issue but could not resolve it.
Meanwhile I was curious to see the 31 rays in three dimensions.
I guessed they were a subset of the same cubic lattice and I did
the search with a computer program. The algorithm appears in my
book \cite{qt}, where there is also an exercise: write a computer
program for the four-dimensional case, and check that 24 rays are
the minimum number. I never bothered to solve that boring exercise,
but two students took up the challenge and found that it was possible
to remove {\it any one\/} of the 24 rays, and still have a KS set.
Michael Kernaghan, in Canada, found a KS set with 20 rays \cite{20}
and then Ad\'an Cabello, together with Jos\'e Manuel Estebaranz
and Guillermo Garc\'{\i}a Alcaine in Madrid, found a set of 18 rays
\cite{18}. They still hold the world record (probably for ever).

Soon after I found my $33+24$ rays, there was the Gulf war and Saddam
Hussein sent numerous scud missiles on Israeli targets. Each time there
was a raid alert, we had to go into an air-tight room and don gas masks
(fortunately they were never needed). Each alert lasted about half an
hour, until the scud debris were examined by civil defense experts. To
help time pass, I read {\it Boojums\/} \cite{boojums} through the
goggles of the mask. This was sometimes difficult, because the goggles
were fogged by my breath, but this was always enjoyable and instructive.

It is a pleasure to dedicate this article to David Mermin, for his
birthday  and many more birthdays.\bigskip

\noindent{\bf2. WHAT CAN BE OBSERVED?}\bigskip

Standard texbooks on quantum mechanics tell you that observable
quantities are represented by Hermitian operators, their possible
values are the eigenvalues of these operators, and that the probability
of detecting eigenvalue $\lambda_n$, corresponding to eigenvector
$u_n$, is $|\6u_n|\psi\9|^2$, where $\psi$ is the (pure) state of the
quantum system that is observed.  With a bit more sophistication to
include mixed states, the probability can be written in a general way
$\6u_n|\rho|u_n\9$. Really bad books also claim that the state of the
physical system after the measurement collapses into the corresponding
$u_n$. This is sheer nonsense. (Finding appropriate references is
left as an exercise for the reader.)

The simple and obvious truth is that quantum phenomena {\it do not\/}
occur in a Hilbert space.  They occur in a laboratory. If you visit a
real laboratory, you will never find there Hermitian operators. All
you can see are emitters (lasers, ion guns, synchrotrons and the
like) and detectors. The experimenter controls the emission process
and observes detection events. The theorist's problem is to predict
the probability of response of this or that detector, for a given
emission procedure. Quantum mechanics tells us that whatever comes from
the emitter is represented by a state $\rho$ (a positive operator,
usually normalized to 1). Detectors are represented by positive
operators $E_\mu$, where $\mu$ is an arbitrary label whose sole role
is to identify the detector. The probability that detector $\mu$ be
excited is $\tr(\rho E_\mu)$. A complete set of $E_\mu$, including the
possibility of no detection, sums up to the unit matrix and is called
a {\it positive operator valued measure\/} (POVM)~\cite{qt}.

The various $E_\mu$ do not in general commute, and therefore  a
detection event does not correspond to what is commonly called the
``measurement of an observable.'' Still, the activation of a particular
detector is a macroscopic, objective phenomenon. There is no uncertainty
as to which detector actually clicked.\bigskip

\noindent{\bf3. IMPRECISE MEASUREMENTS} \bigskip

\noindent There has recently been considerable controversy about the
possibility of testing the physical implications of the KS theorem or
even ``nullifying'' the latter \cite{mkc}. The ``observables'' that are
usually considered are spin components of a spin-1 particle, which is
the paradigm of the KS theorem. Opinions are varied \cite{other} but
more often than not assume that the result of an imprecise measurement
of a spin component ${\bf n\cdot J}$ is the value of another spin
component, along some direction ${\bf n'}$ which is close to the
correct {\bf n}. As shown below, this assumption is generally unfounded
(except in the trivial case of spin-$1\over2$ particles).

When I first learnt of this ``nullification,'' my gut feeling was that 
the claim was not even wrong. However, finding a decisive argument
that would convince the authors \cite{mkc} proved as elusive as giving an
experimental proof of a violation of Bell's inequality. It is always
possible to find loopholes. All I can do is to challenge them to find a
loophole in the following argument.

Whatever the causes of imprecision are, if we don't measure exactly
${\bf n\cdot J}$, we have to give a physical interpretation to the 
result of our experimental procedure.  Do we measure some other nearby
operator,
$$ A={\bf n\cdot J}+\sum_{m,n} a_{mn}\,J_m\,J_n/\hbar, $$
where the unknown dimensionless coefficients $a_{mn}=a_{nm}^*$
are small? If the array $a_{mn}$ is antisymmetric, and therefore
imaginary, we have $A={\bf m\cdot J}$, where ${\bf m\approx n}$ (but
{\bf m} need not be a unit vector). However, in general, the operator
$A$ defined above also contains bilinear terms which cannot be reduced
to another component of {\bf J}. These terms have no classical analogue
and no name in our vocabulary. Quantum mechanics has a much richer
choice of dynamical variables than classical mechanics.

Actually, the result of the measurement procedure may not even be
the value of any dynamical variable that looks like $A$. Traditional
concepts such as ``measuring Hermitian operators,'' that were borrowed
or adapted from classical physics, are not appropriate in the quantum
world. In the latter, as explained above, we have emitters and
detectors, and calculations are performed by means of POVMs.

Any lack of precision in the experimental setup simply means that
the positive operator $E_\mu$ which is actually implemented in the
laboratory only approximates the original $E_\mu$ that the experimenter
intended to use. That $E_\mu$ may have been associated with a classical
interpretation such as the value of a component of angular momentum.
The actual $E_\mu$ may have no classical interpretation at all. In
particular, there is no reason to expect it to correspond to a component
of {\bf J} along a slightly different direction.

What does this imply for attempts to test experimentally the KS
theorem? That theorem is only a statement about Euclidean geometry. Any
purported experimental test should be analyzed as explained above:
particles are emitted according to a specified procedure, and quantum
theory is used to predict the probability of excitation of various
detectors. After we take into account known imperfections of the
emission and detection processes, any discrepancy would imply either
a poor understanding of the laboratory equipment, or a failure of
quantum theory. The KS theorem itself is not involved. Its only role
is to restrict our freedom of concocting realistic non-contextual
theories that would replace quantum mechanics.

\bigskip\noindent{\bf Acknowledgment}

\bigskip\noindent This work was supported by the Gerard Swope Fund
and the Fund for Encouragement of Research.

 \end{document}